\documentclass{elsart5pag}

\usepackage{natbib}

\usepackage{graphicx}

\usepackage{amssymb}

\begin{document}

\begin{frontmatter}



\title{ More than two equally probable variants of signal in Kauffman networks as an important overlooked case, negative feedbacks allow life in chaos}


\author{Andrzej Gecow\thanksref{homead}}
\ead{gecow@op.pl}
\thanks[homead]{01-923 Warsaw, Boguslawskiego 4/76, Poland, gecow@op.pl}
\address{Centre for Ecological Research Polish Academy of Sciences,\\
M. Konopnickiej 1, Dziekanow Lesny, 05-092 Lomianki, Poland}

\begin{abstract}

There are three main aims of this paper. 1- I explain reasons why I await life to lie significantly deeper in chaos than Kauffman approach does, however still in boundary area near `the edge of chaos and order'. The role of negative feedbacks in stability of living objects is main of those reasons. In Kauffman's approach regulation using negative feedbacks is not considered sufficiently, e.g. in gene regulatory model based on Boolean networks, which indicates therefore not proper source of stability. 
Large damage avalanche is available only in chaotic phase. It models death in all living objects necessary for Darwinian elimination. 
It is the first step of my approach leading to structural tendencies which are effects of adaptive evolution of dynamic complex (maturely chaotic) networks. 
2- Introduction of $s \geq 2$ equally probable variants of signal (state of node in Kauffman network) as interpretively based new statistical mechanism (RSN) instead of the bias $p$ - probability of one of signal variants used in RBN family and RNS. It is also different than RWN model. For this mechanism which can be treated as very frequent, ordered phase occurs only in exceptional cases but for this approach the chaotic phase is investigated.
Annealed approximation expectations and simulations of damage spreading for different network types (similar to CRBN, FSRBN and EFRBN but with $s \geq 2$) are described. Degree of order in chaotic phase in dependency of network parameters and type is discussed. By using such order life evolve.
3- A simplified algorithm called `reversed-annealed' for statistical simulation of damage spreading is described. It is used for simulations presented in this and next papers describing my approach.

\end{abstract}

\begin{keyword}
Boolean network \sep damage spreading \sep chaos \sep living system \sep algorithm.

\end{keyword}
\end{frontmatter}

\section{ Introduction - Differences with Kauffman's approach }
\label{ch.1}

The Kauffman's approach \citep{Kauf69,Derrida86,Kauf90,ooKauf,Luque05,Iguchi07}  to description of living objects (or a system designed by human) using dynamical directed networks is now the most attractive and promising idea, currently developed by many authors. Reading his `The Origins of Order' \citep{ooKauf}  I was highly excited and full of admiration. However, my understanding of few basic assumptions and conclusions is different and in those areas I develop a modified approach.

\subsection{ Contents of the paper in brief }
\label{ch.1.1} 

A. I estimate that living objects or a system designed by human are purely chaotic, however still in `liquid region' - that is area near `the edge of chaos and order' as suggested by Kauffman. My estimation is based mainly on: \\
1- simple observation of damage spreading in reality  when initial change is not of `known, prevented type' (typically - damage avalanche possible only in chaotic phase and interpreted as death) (ch.1.3.2);\\
2- lack of negative feedbacks in adequate level in Kauffman's investigation of stability especially of gene regulatory model (ch.1.2.2) which makes Kauffman's approach expectation of `life on the edge of order' based on stability less convincing;\\
3- finding that $s$ - number of equally probable signal variants typically should be greater than two which leads to chaos (ch.1.5); \\
4- remarking that the ordered case ($s=2,K=2$) is an extreme exception in the `sensible' (without fixed $K=1$) network parameter space (ch.1.5.4, ch.2.1);\\

B. The main aim of this paper is to introduce \citep{paris,hof,Luque97,Luque04,Luque05} more than two equally probable signal variants ($s\geq 2$) to the Kauffman networks instead of only two variants in Boolean networks with bias $p$ differing their probability (thus expanding Random Boolean Network RBN to Random Signal Network RSN). The reasons are as follows:\\
1- It is an important overlooked case which has a good interpretative base - it is probably typical case (second variant contains all remaining real but not interesting variants) ch.1.5.2. 
Equal probability of variants is a natural and useful first approximation which allows for wide investigation, e.g. to expand the expectation made for $s=2$ in annealed approximation by 
\citet{Derrida86} also for $s>2$  (ch.2.2, fig.1). \\
The statistical mechanisms for such case are different than the ones described using bias $p$ in the classic Boolean case and lead to different results ch.1.5.3.\\
2- It is also different than sister networks using more signal variants than two: 
RWN (Random Walk Network) construction \citep{Luque04, Luque05} which contains memory in its nodes' states or 
RNS \citep{Luque97, Sole00} (Random Network with multiple States) where bias $p$ is used (for one of signal variants, making it different from the other ones) for reaching edge of chaos. \\
3- Boolean networks are not generally adequate. They can describe all cases but the requirement of exactly two signal variants introduces not real states or other discrepancies to described reality ch.1.5.1.\\

In this paper I am going to show that such a solution can be better and lead to more adequate description and predictions. Interpretation is the base of the proposed assumption $s>2$, but I also show that using it in the Derrida `annealed approximation' and the `semi-quenched' model (named here `reversed-annealed' ch.3.2) significantly different results are obtained (ch.2 fig.1 and ch.3.3 fig.6).

C. The `reversed-annealed' algorithm (described in ch.3.2) used for simulation is a special simplified algorithm dedicated for statistical investigation of damage spreading in Kauffman and similar networks in synchronous mode. It is not an optimisation algorithm but it can be used to study an adaptation process. The main properties of this algorithm are:\\
1- Only damaged nodes are calculated here. \\
2- The process stops when damage fades out or achieves the equilibrium level. \\
3- Therefore it is much more optimal algorithm. In the classic method two full systems are calculated - the disturbed one and the undisturbed one and at each time step they are compared to measure the damage. \\
4- This simplification has small meaning for investigations presented in this paper, however, it is a critical key to the area of `structural tendencies' in adaptive evolution (shortly described in   \citep{dgec}) which are the aim of the path starting with this paper. See also \citep{bgec}.\\ 
5- The main feature of this algorithm is that it gives a single particular output state, however only statistically correct, instead of long cyclic attractors. In my approach avoiding attractors is useful in opposition to Kauffman's approach where properties of attractors are very interesting.\\

In ch.2 theoretical consequences of expansion to $s\geq 2$ are discussed. A simple and intuitive `coefficient of damage propagation' $w$ is introduced and the Derrida's annealed approximation is expanded.

In ch.3 influence of parameter $s$ on networks features is studied. Simulation and its algorithm for different network types are described.

\subsection{Spontaneous order, negative feedbacks}
\label{ch.1.2} 

\subsubsection{ Is the `spontaneous order' a real choice of order?}
\label{ch.1.2.1}

One of two basic theses of Kauffman's approach is that `spontaneous order' is a significant part of observed order in living objects. I can agree but what is `spontaneous order'?  
(See Appendix B.) 
When we are going to understand living objects, it makes sense to deal with inanimate objects or systems first. We  can define our currently expected `abiotic' set of possibilities  - i.e. set of possible objects and their approximated distribution of probability within inanimate objects. 
Next, we compare it to the observed reality (also inanimate objects only) and we encounter certain differences, e.g. observed objects cover only a small part of our set. This discrepancy is the observed `spontaneous order'. It is here a subjective phenomenon, not a real one. It depends on our expectation which turns out to be inadequate.

Let's consider an example\footnote{It is the original example which Kauffman used (p.201) to explain a base of his hypothesis about spontaneous order. I discuss here this base. Later finding of Samuelsson and Troein (Phys. Rev. Lett. 90, 098701 (2003)) that used here the expected median state cycle length scales faster than any power law, changes only parameters of this base making them less extreme, but doesn't change the main view. Approximated number like `317'  will change significantly but nor regarding this problem.} from Ref\citep{ooKauf}. For Boolean network which Kauffman investigated and $K=2$ inputs per node (ordered case) the expected median state cycle length and the number of state cycle attractors is $N^{1/2}$. It means that a system of size  $N=$100 000 (comparable to the human genome) would have about only 317  alternative asymptotic attractors which use similar number of states, but number of available states is $2^N$. It is spontaneous order that is shown there. Why we expect so many possibilities ($2^N$) if we know that there can only be so few (317)? It is property of any random Boolean network with $K=2$, not only a living one. For other low $K>2$ this `spontaneous order' is also very significant, however, not so extreme.  

I do not suggest that a snow star, the beautiful fractal and symmetrical structure, is not really ordered spontaneously, but what is the real set of possible shapes from which this spontaneous order makes its choice? Does it contain a square? If it contains only possible shapes of snow stars, then there is no choice between ordered or not ordered shapes. What is a difference if we find a metallic spur gear with 6 teeth? Each time when we meet a visible pattern, then we are looking for a goal. For the snow star there is no goal, there is a mechanism of its creation only, but for spur gear a goal is essential - it is the base of creation mechanism.

Now we observe living (adapted) objects with certain low $K$, we also encounter the same `spontaneous order' but we find a more specific distribution.
The remaining part of the observed order is a real one, i.e. it is independent of our expectations. It is the discrepancy between the abiotic state and the observed one. 
This part should be treated as `purposeful information'\citep{R1,R1rgec} 
or `biological information'\citep{Kuppers} and Darwinian selection is its only source as stated by Kauffman. The problem is: which observed order (pattern) is the purposeful information, i.e. source of stability, because stability is a goal of living object. An overestimation of spontaneous order and connected with it homeostatic and structural stability (small initial change gives small effected change, see Appendix B) leads to type of creationism without God, i.e. certain purposeful information is taken as created (spontaneously) without natural mechanism which can create such information. (Purposeful information is a choice of cause which leads to assumed effect, i.e. goal. Only for the goal: `continuation of existence' such purposeful information can be spontaneously collected using Darwinian mechanism \citep{R1,R1rgec}.)

Note, in such an approach there is by definition no place for real `spontaneous order'.
However, reality of `spontaneous order' shouldn't be accepted if we can find such reality (as above) to compare (without living objects) and we cannot continue treating our large set as a real abiotic one. For complex dynamical networks it can be a problem to find such reality and someone can state that abiotic set doesn't contain such complex networks. If so, then the same, earlier subjective phenomena stay similar to real `biological information' as is in the Kauffman's view. 
Note, however, that order which is connected with `biological information' is connected also with `purposeful information' \citep{R1,R1rgec} but in the case of `spontaneous order' such connection doesn't exist which has basic interpretative importance.
This change of point of view is only allowed as long as we don't find real objects described by complex random Boolean network which are not living ones. Snow stars exist, however, they probably cannot be described so simply.
I incline to the view that complex networks created without adaptive condition do not have to describe living objects only. (In the life process I include whole human activity and human products.)

\subsubsection{ Is the `ultrastability' considered in Kauffman approach?}
\label{ch.1.2.2}

To keep a specific state, i.e. not a typical spontaneous state, there are necessary some `stability' mechanisms which block spontaneous transitions to a more probable spontaneous state. The biological information is such a specific state, it can be lose and therefore it needs such a stability mechanisms. These mechanisms typically are some regulators based on negative feedbacks. Creation of these mechanisms is an effect of Darwinian selection. It is commonly observed as extremely high concentration of negative feedbacks in living objects which is even used for life definition \citep{Benio01, Benio05} as its specific property.

In Kauffman approach such `stability' mechanisms are named `ultrastability' and are taken as one of base. They are, however, considered only on random normal average level of appearance without any preferences, which is a very large simplification leading to non-realistic modelling of stability of living objects. 

Following 
\citet{Ashby}, Kauffman uses `essential variable' to describe ultrastability: ``In the context of Boolean networks, keeping the essential variables in bounds corresponds most simply to holding them fixed'' (page 211 in   \citep{ooKauf}). Later Kauffman looks for systems with `frozen' areas and finds them near the phase transition. In the ordered phase frozen areas percolate leaving inside small isolated lakes of activity which leads to `homeostatic stability'. In the chaotic phase there are isolated islands of frozen areas.

I have checked how the above assumption works using simple known example of thermostat in fridge.
The details can be found in Appendix A. 
In conclusion I state that assumption of fixed states for essential variables is a too strong simplification for description of negative feedbacks. In effect which occurs opposite to intention, negative feedbacks are removed out of such a network. 
This way `homeostatic stability' based on `spontaneous order', i.e. typically `structural stability', is taken as an explanation of stability of gene regulatory network \citep{Kauf71,Wagner01,Serra04,Shmul05,Ramo06,Serrajtb07,Serra10}  instead of the neglected homeostasis based on negative feedbacks. This is the above mentioned (ch.1.2.1) overestimation of spontaneous order. As can be found in   \citep{RekAlb03}  as conclusion of real gene network investigation, ``steady states are determined by the topology of the network and the type of regulatory interactions between components'', however, the term `feedbacks' is not yet used there.

Kauffman's gene regulatory model\citep{Kauf71} 
or even\citep{Kauf69} 
is still alive and developed\citep{Serra10,Serrajtb07,Sole00,Luque05} mainly because of its high level of abstraction. Its results gave new perspective which bear fruit of other gene regulatory models, e.g. more connected with genetic reality (less abstractive) GRN's which are developed basing on 
\citet{ Banzhaf03} work. 
Evolution of such networks is based on Darwinian mechanisms e.g.  \citep{Knabe06} where biological internal clock emerges. Such simulations together with real network mapping gives base to measure of fraction of negative feedbacks which is crucial to define proper source of stability.

\subsection{Estimating living objects as chaotic}
\label{ch.1.3}

\subsubsection{ `Not fully random' - `known' changes }
\label{ch.1.3.1}

Homeostasis built by Darwinian selection and based typically on negative feedbacks leads to stability for certain set of initial changes. Such changes may randomly occur but 
they are `known' to system which means that the system reacts to them in a non-random way. 
The remaining changes can be treated as `fully random' and when they occur, the system can behave in a chaotic or ordered way. 
We should focus our investigation on this set of `fully random' changes but considering stability of system we cannot neglect `known' changes.

Current living objects `know' most of typical changes (in their typical environment) and therefore exhibit high stability even though for fully random (unknown) changes they are chaotic.
Similar idea of splitting network body by removing certain special part (frozen) which covers the feature of normal rest was applied in   \citep{Bilke01} in `decimation algorithm' for frozen nodes in ordered phase. \citet{ Bornholdt00} tends in opposite direction (preferring such nodes) but less openly.  It cannot be implemented to prefer the later discussed part of small changes which are selected only from peak of ordered behaviour because these peaks are not parts of the network (they are formed by change events, not network nodes). 

Which changes are `fully random' (not predicted in the system structure)? For systems designed by human a designer knows the answer but for living objects it is a problem. We can expect that surgical operation on the brain or heart, or strange environment influences are not predicted in our structure as living objects. Small defects in DNA copying are predicted and certain set of repair or other safety mechanisms are prepared - it is predicted that a defect will occur even though any specific defect is unpredicted. Conditional specification characteristic of all vertebrates can neutralize lot of different events which seem to be random but therefore they are not (fully) random. What range of effected change should we typically expect when particular initial change is random and unpredicted in the above sense? Can we expect homeostatic stability based on spontaneous order or structural stability (see Appendix B) as was the case for ordered systems? I believe that there can be only one answer.   
If the answer `yes' is considered, then regeneration of body plan in lower animals is probably invoked. Note in such a case that evolution to later `higher' animals lost this feature, but following Kauffman's idea it should rather have neared to edge of chaos and order (see ch.1.4) and intensified such feature. 

\subsubsection{ Large damage avalanche modelling death typical for `fully random' changes is available only in chaotic system }
\label{ch.1.3.2}

High stability of ordered systems does not allow for damage to evolve into a large avalanche. (See Appendix B.)
The typical behaviour of damage spreading for fully random (not predicted in system designing  process) initial change is the main criterion of my estimation that the typical living object or a system designed by human are chaotic. I estimate that such initiation causes with high probability a large avalanche of damage which may be observed only in chaotic systems. 
This estimation, however, is only intuitive, I will not prove it in this paper but an investigation can be made to verify it.

What can be the interpretation of large damage avalanche? - It can be only death and elimination. 
Any living object can die. It is common important feature of living objects necessary for Darwinian elimination which creates purposeful (biological) information. It must be an element of the model. 

Life is a continuous maintenance of equilibrium at a high level, which is a semi-unstable equilibrium - it  will collapse into a large damage avalanche after a single false move. This view correctly describes chaotic state but not an ordered one with its `homeostatic stability' implied from `spontaneous order'. Therefore observed stability cannot be spontaneous - means `ordered' but carefully collected which only Darwinian selection can find and not in form suggested by Kauffman (place near ordered phase) but in form active regulation typically - negative feedbacks. 

In the ordered phase there is no such radically different possibility which can model death. (See fig.8 in Appendix B.)
Large damage avalanche cannot be a transition to a different acceptable attractor.  It is such a large change of function, that most of the mechanisms of homeostasis based on negative feedbacks (biological information) lose their parameters needed to function and cannot work. Such `biological information' disappears. Natural criterion of identity \citep{R1,R1rgec} based on small change defines such changed system as not `the same'. Note, we discuss initial  and effective change in system `structure and function' space, not in the fitness space, however, change in the `structure and function' typically causes a change in fitness.

Living objects are self-maintained systems. After damage avalanche in simulated network we still have the same system (structure doesn't change) but in other basin of attraction. We, however, should remember that large change of living object in function cause large change in maintenance and in effect large change in system structure. It may be interpreted that part of node states describe structure of evolving network. After death structure of earlier living object drastically change and its function is also drastically different. It becomes absolutely other system.

After the above discussion, let us come back to the \citet{Hughes00} experiment analysed in  \citep{Wagner01,Serra04,Ramo06,Serrajtb07,Serra10} using Kauffman gene regulatory model. This analysis is considered as the main argument behind the `life on the edge of chaos' hypothesis. In this experiment only those cases with knocked out gene are analysed which grow after this disturbance (that is, which are still alive). There are no cases of death in this set. It means, that all measured effects of disturbances are within the range of homeostatic answer. In this range living systems indeed behave like ordered ones with frozen essential variable, which is not suprising.

\subsection{ Living objects there are on opposite shore of liquid region near edge of chaos }
\label{ch.1.4}

The area between chaotic and ordered phase exhibits highest `structural stability' (small initial change gives small effected change) which is useful for adaptive evolution. Maximum of spontaneous order also occurs there. Kauffman isolates this area as third region named `liquid'.
Evolution using random walk in the space of network parameters should tend to this area because there effects of evolution are larger. This expectation of Kauffman's approach is known as `life on the edge of chaos'. Kauffman expects of living systems rather on ordered shore. (See Appendix B.)

I can agree that evolution tends to this area because of structural stability, however, I am afraid that nearing to the edge of chaos is hard to reach and life is still far from this `optimal' (with respect to the single property of structural stability in this simple model) area. I estimate above that living objects are chaotic. However, in chaotic even large networks a part of initial changes doesn't effect in chaotic `damage avalanche' but does effect in small change available for evolution. Fraction of such small changes is higher if parameters of network are nearer phase transition between order and chaos. This view can be included into Kauffman's idea of `liquid region near phase transition', however, living systems exist there on opposite shore. This theme is developed in this paper in ch.3.3 where in simulation results in fig.6 parameter $r$ is discussed.

It is interesting what specific properties these small effective changes in chaotic regime have. 
I pick up this question in my research which leads to `structural tendencies' \citep{dgec}. They are the main goal of my  approach which allows to explain old classic regularities in ontogeny evolution which were still waiting for explanation.
Structural tendencies bring us back to the base of Kauffman's ideas of structural stability and life on the edge of chaos. This base is the obvious condition of small changes constituting adaptive evolution. 
I call it `small change tendency' \citep{paris,hof,krab,dgec}. In ch.3 and in conclusion of this paper we find that the size of effected change initialised in chaotic systems has a distribution with two peaks - for large changes (equilibrium level of damage avalanche in circular attractors) and for very small changes which are an effect of real fade-out of damage in first few steps. The real fadeout is connected to ordered behaviour, damage avalanche (pseudo-fadeout in my algorithm) - to chaotic behaviour. In the next papers on my approach it will be found (see also \citep{dgec}) that small change tendency selects for adaptive evolution only small changes from the peak of ordered behaviour (as Kauffman correctly expected), but chaotic systems do evolve. 

Definitions of particular structural tendencies needs a certain frame of reference in network body. The network external outputs are good such frame, especially when fitness is defined on them which is in my opinion more adequate than the suggestion made in Boolean NK model \citep{ooKauf} where fitness is a fraction of nodes with `proper' state.
The path to results of simulations of structural tendencies isn't short and it exceeds the frame of this paper (which is only the first step on this path). In this paper we will not yet consider fitness nor external outputs but an extended base (B. i.e. $s>2$ and C. `reversed-annealed algorithm') for investigation of chaotic (A.) network will be defined.

\subsection{More than two equally probable signal variants}
\label{ch.1.5}

\subsubsection{ Boolean networks are not generally adequate }
\label{ch.1.5.1}

As it implies from example in Appendix A Boolean network appears not adequate to precise description of a simple typical thermostat. In reality the temperature is split into three sections $a,b,c$. In one time point a temperature can be only inside one of them. Using three variants signal (state) is natural and leads to simple, correct network (fig.7.2), but requirement to use only two variants (alternatives) leads to losing an important aspect in description (fitg.7.1) or to creation of dummy states (fig.7.3).

\subsubsection{ Two variants  are often subjective }
\label{ch.1.5.2}

Typically while describing adaptive systems  we encounter two alternatives but they usually have very different probabilities. (Typically parameter $p$  called `bias' was used as the probability of one of these alternatives.) It can be, however, defined in a very subjective way, I have a certain explanation how it can happen: 
One of the typical ways leading to two alternatives is our concentration on one particular event and collecting all the remaining events as the second alternative. This second alternative is NOT the first one, and we obtain two alternatives. There are lots of alternatives in such a case in the reality and this is important for the statistical mechanism but we are only interested in one of these alternatives. Typically it is `the proper event' when we are concerned with systems which adapt. 
Note, for system which adapt the notions: `proper' and `correct' are defined using fitness but it has nothing to do with the statistical mechanism and such simplification remains subjective.
This is the main, however, simple and important cause of introducing more than two alternatives. 
Note, I estimate that this cause of observation of two alternatives is typical.

If we are going to describe e.g. the long process leading from gene mutations to certain properties assessed directly using fitness, then more than two alternatives for the description of mechanism of such a case seems much more adequate.  We should remark that there are 4 nucleotides, 20 amino acids and other unclear spectra of similarly probable alternatives. In this set of spectra of alternatives, the case of as few as two alternatives seem to be an exception,
however, for gene regulatory network it seems to be adequate in the first approximation (active or inactive gene). 
Investigators of real gene networks suggest: ``While the segment polarity gene network was successfully modelled by a simple synchronous binary Boolean model, other networks might require more detailed models incorporating asynchronous updating and/or multi-level variables (especially relevant for systems incorporating long-range diffusion).''\citep{RekAlb03} 
In second approximations which are RNS \citep{Luque97, Sole00}  and RWN \citep{Luque04,Luque05} models with more than two variants are used but in a different way than here (RSN). 

\subsubsection{ Equal probability of $s\geq 2$ variants of signal as alternative to bias $p$ and RWN }
\label{ch.1.5.3}

Using equal probability of alternatives is probably the only way to define from within the model a probability necessary for prediction and calculation. In such a way we obtain $s$ (which can be more than two) equally probable signal variants ($s \geq 2$) which I am going to introduce in this paper into the Kauffman networks used for general description of real adaptive systems. 

Parameter $s$ describes an alternative (XOR) new mechanism other than the bias $p$ - probability of one of two alternatives (or first one of more than two in RNS \citep{Luque97,Sole00}). 
For $s>2$ (and $K \geq 2$) damage should always statistically grow (whenever it has room to grow) which my `coefficient of damage propagation' shows, and we should always obtain chaos. 
For extreme bias $p$ and small $K>2$, however, we obtain order \citep{Derrida86,Aldana03,Fronczak08}.

Similar to bias $p$ is $P$ - the `internal homogeneity' in Boolean functions \citep{ooKauf}. 
Note that parameters $s$ and $P$ lie into opposite direction when they differ to their typical, smallest value. Higher $s$ causes chaos but higher $P$ allows to avoid chaos, however both of them are connected to similar problem of equal probability of two variants of signal. They describe different aspects of this idealisation. 
An other deviation of random function drawing can be found in the literature. It may have some connection to considered theme of signal diversity. 
A Boolean function is said to be canalising if at least one value of one of it's inputs uniquely determines it's output, no matter what the other input values are. Real gene networks show a proportion of canalising functions much higher than the one corresponding to a uniform choice so it is used the set of canalising Boolean functions only (i.e. for K=2 without XOR and NOT XOR) \citep{Harris02, Serra04}.
 
In RNS these signal variants are not equally probable. There bias $p$ plays an important role allowing investigation of phase transition to chaos like in the whole Kauffman's approach. It is not a mechanism substituting bias $p$ although using $p=1/s$ RNS formally contains my RSN. In my earlier publication \citep{paris,hof}, however, more than two signal variants also can be found and they are equally probable like here and like in a small paragraph in   \citep{3Aldana}. 
Typically the case of more than two variants which is taken as interpretatively better \citep{3Aldana}, is rejected \citep{3Aldana} or not developed as contradictory with the expectation of `life at the edge of chaos' which we reject here. 
Construction of RWN also is an effect of this condition. Its functions perform a shift of node state and this way create memory. This shifting reaches upper or lower boundaries and the result is random but in a complicated way. 
(All networks discussed in this paper are deterministic in their function, randomness is used to define the construction and the beginning state. The annealed approximation model concerns  probabilistic automata \citep{Derrida86W}, it is not deterministic. )  
Mechanisms described using RWN are different than ours, they have to be described using differential equations. It is a step towards regulators and negative feedbacks, however, in this model homeostatic stability based on negative feedbacks has no pressure to emerge. 

\citet{Sousa05} considers the scale-free network and more than two different opinions and he obtains a vote distribution  which is in better agreement with reality. Similarly 
 \citep{Stauffer2004,Jaco2005,Stauffer2006} consider $Q$ opinion states. 

\subsubsection{ `Coefficient of damage propagation' simply shows that case $s=2$ is extreme }
\label{ch.1.5.4}

As typically in literature, let $K$ be the number of node inputs and $k$ - the number of node outputs (outgoing links). I assume constant $K$ i.e. equal for all nodes of the particular network as in early 
 \citet{Kauf69,ooKauf} works. I neglect the extreme strange case of $K=1$ in my investigation of network features (however in example in Appendix A it appears - fig.7.1 and we can find the case $K=1$ in the literature \citep{ooKauf,Wagner01} and $1 < K <2$ in newer works  \citep{Iguchi07} where $K$ is flexible for particular network) therefore $K$ has to be greater or equal $2$. We consider an autonomous network therefore $\langle k\rangle=K$ and $\langle k\rangle\geq 2$ but for a particular node $k=1$ and $k=0$ can happen. 
For completeness I repeat: $s$ is the number of equally probable variants of signal. Note, that using $s$ we know that they are equally probable. Bias $p$ is replaced by this new parameter $s$.

Up till now the term `Kauffman network' was synonymous with `Boolean network', but not anymore. The term `Boolean' must be limited to two variants of signal but `Kauffman network' can and should contain more signal variants as is used here (RSN) or in RWN or in RNS models.

I introduce \citep{paris,hof,3Aldana} (ch.2) a simple intuitive indicator of the ability of damage to explode. This is $w=\langle k\rangle(s-1)/s$  `coefficient of damage propagation' (or `damage multiplication' on one element of system if only one input signal is changed) which indicates how many output signals of a node will be changed on the average if one input signal is changed (for the random function used by nodes to calculate outputs from the inputs). (I assume minimal $P$ - internal homogeneity \citep{ooKauf} for the whole of this paper and approach.) 
In the   \citep{3Aldana} similar equation (6.2):  $K_c = s /(s- 1)$ is given for condition $w=1$ i.e. to keep the system within the ordered phase. I use it for an opposite purpose.  $K_c$ is here the critical connectivity which for the minimal $s=2$ takes its maximal value of 2. 

Coefficient $w$ is interesting for the whole network or for its part, not for a single particular node. However, it is easier to discuss it on a single, averaged node. Therefore I have started my approach using aggregate of automata \citep{paris,hof,krab} (ch.2.3) where $K=k$ and each outgoing link of node has its own signal. It is different than in case of Kauffman network where all outgoing links transmit the same signal. 
Neglecting $K=1$ for the whole network  the value $w\leq 1$ occurs only in case of $s=2$ and $K=2$.  For all other cases there is $w>1$ and a small initial damage should statistically explode onto a large part of the system, which means that the system is chaotic. 
For a part of network the average node degree $k$ can be less than two and locally there can be $w<1$. For definition of such network part, distance to external output of the network can be used (this theme, however, will be discussed in another papers, see also   \citep{dgec,bgec}).

The case $s=2$ is extreme - there is no smaller sensible (natural) value for $s$ - but it is even more extreme than we have seen. If we simultaneously use $\langle k\rangle=2$ in the same way, then we obtain an especially extreme case - order instead of chaos. It is the case $K=2$ for autonomous Boolean networks which is known \citep{Derrida86,ooKauf} as the critical one.  Such extreme values of parameters assumed for the model should have special known causes, other than useful simplicity of the model, especially if we need chaos as we estimate above. Such a special cause could be satisfying the assumption of `life on the edge of chaos' (which we reject here). Generally a safer approach is to use not so extreme values for an unknown parameter. 

\subsubsection{ The damage equilibrium levels for $s>2$ are significantly higher }
\label{ch.1.5.5}

Number $s$ of equally probable variants of signals is the next main parameter of system, like bias $p$, Kauffman's $K$ - number of  inputs per element and $P$ - the internal homogeneity in Boolean functions. These parameters define a system as chaotic or ordered. 

Next we examine the differences of effects obtained using $s \geq 2$  and the $s = 2$ only. 
I find two important differences.

I find the first of them by expanding the Derrida \citep{Derrida86,ooKauf} method (annealed approximation) of calculation of equilibrium level of damage (ch.2.2) to cases of $s>2$. The levels for $s>2$ are very different from the case of $s=2$ (see fig.1). The parameter $s$ has a much stronger influence on these levels (moving up their upper limit two times) than the $K$ parameter used up till now for exploration of chaotic regime, therefore $s>2$ cannot be neglected and substituted by $K>2$ for this exploration. 

\subsubsection{ Importance of parameter $s$ from simulation }
\label{ch.1.5.6}

The result of simulation shows the second important difference (ch.3.3). The parameter $s$, especially for lower values, has a significant influence on the behaviour of different network types (see fig.6), especially scale-free networks, in the first crucial period of damage growth.

Both these differences in effects of the assumptions $s=2$ and $s>2$ (even neglecting the possibility of phase transition from chaos to order) enter the range of qualitative differences. This confirms the importance of this choice. 

The assumptions of two variants and their equal probability are also used in a wide range of similar models like e.g. cellular automata, Ising model or spin glasses \citep{Jan94}.  They are typically applied as safe, useful simplifications which should be used for preliminary recognition. But just like in the case of Boolean networks these assumptions may not be so safe and should be checked carefully. In  the original application of Ising model and spin glasses to physical spin they are obviously correct, but these models are nowadays applied to a wide range of problems, from social (e.g. opinion formation \citep{Kos06}) to biological ones,  where such assumptions are typically simplifications.

\section{Expectations for $s>2$ - Coefficient $w$ of Damage Propagation and Annealed Approximation}
\label{ch.2}

\subsection{Definition and meaning}
\label{ch.2.1}

Let us define a coefficient $w$ of damage propagation: \\  
$w=\langle k\rangle(s-1)/s$ where $k$ is the number of node outputs. Kauffman in the first step considers 
constant \citep{Kauf69,ooKauf} (for a particular 
network) number $K$ of node inputs, and considers an autonomous network which 
has no external inputs or outputs. For autonomous networks $\langle k\rangle=K$: 
the average  $k$  in the network is equal to the
fixed $K$. The coefficient $w$ shows how many output signals of a node
are changed on the average if at least one of its input signals is
changed. In the case of  `one changed input signal' this
coefficient can be named `coefficient of damage multiplication
on one node'. If it is greater than one ($w>1$) then damage should
statistically grow and create an avalanche which spreads onto a
large part of a system. It is similar (especially in the first, crucial period of a process) to the coefficient of neutron
multiplication in a nuclear chain reaction - if it is less than
one then we have a nuclear power station, if it is greater than one
then an atomic bomb explodes. 

The coefficient of damage multiplication depends on the functions draw
- if functions are not properly random then the coefficient $w$ may be
greater or less than the above. The coefficient of damage
multiplication is a simple and intuitive indicator of the possibility of
damage avalanche and therefore of the system's place on the
chaos-order axis but it is only the first approximation as we will
show later. 
Note that case $w\leq 1$ appears only for case  $\langle k\rangle\leq 2$ and
$s=2$ and both these parameters are here in their smallest values.
For all other cases where $s>2$ or $\langle k\rangle >2$ we have $w>1$ and in such a case
damage statistically should explode onto a large part of the system which therefore is chaotic.

The number $s$ of equally probable variants of signals is the next among the 
main parameters of a system which define the system as chaotic or ordered.
It is similar to Kauffman's $K$ - number of node inputs and $P$ - the 
internal homogeneity in Boolean functions or to bias $p$ (- probability of first 
of alternatives) which it substitutes. 
In the first theoretical approximation the coefficient $w$ can substitute two 
of them ($s$ and $K$) in this role but other important features of a system 
depend on the parameters $s$ and $K$ individually and differ although the 
coefficient $w$ is the same.
One of such features is the level of damage equilibrium for chaotic networks which differs much stronger in dependency on the parameter $s$ than on the parameter $K$ (next, ch.2.2). 
The second such features occurs when we investigate various types of networks, differing mainly in the distribution of node degree $k$: we obtain different results in `real fade out' especially for low $s$ and scale-free network than for higher $s$. Such conclusion is an effect of simulation described at the end of this paper (ch.3.3).

\subsection{$w^t$ describes first critical period of damage spreading,
annealed approximation expanded for $s>2$ }
\label{ch.2.2}

\begin{figure}[ht]
\begin{center}
\includegraphics[width=8.5cm]{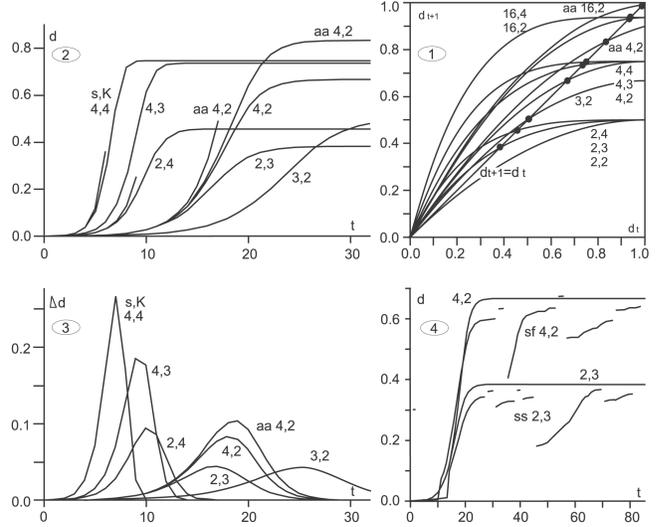}
\end{center}


\caption{Theoretical damage spreading calculated using annealed approximation.
 (1) Damage change at one time step in
synchronous calculation of system known as the `Derrida plot', 
extended for the case $s>2$ and for $aa$ network type. 
The crossing of curves $d_{t+1}(d_t)$
with line $d_{t+1}=d_t$ shows equilibrium levels $dmx$ up to which
damage can grow. These levels are reached in (2) on the left
which shows damage size in time dependency. A simplified
expectation $d(t)=d_0w^t$  using coefficient $w$ is shown (Three
short curves to the left of the longer reaching equilibrium). This
approximation is good for the first critical period when $d$ is still
small. (4) shows examples of experimental curves in comparison
to their theoretical expectations (see ch.3.3. and table 1). In (3)
the increase of damage in consecutive time steps is shown.
Experimental curves are similar but wider, the small difference at
the point of maximum is shown in table 1.}

\label{fig:1}     
\end{figure}

When the avalanche is still small and the range of interaction is
a whole and big system (large number $N$ of elements of system)
then the probability of more than one changed input signal is also
small and damage is well described by $w$ as $d(t)=d_0w^t$  which
is shown in fig.1.2. This is a critical period of time $t$, when
damage is still so small that probability of its fade out is not
to be neglected. Later it practically cannot fade out but the cases of 
more than one changed input signal  occur more and more often and the
real multiplication of damage becomes smaller and smaller up to
the moment of achieving a stable level of damage (fig.1.1 and
fig.1.2). These figures are calculated in a theoretical way 
based on annealed approximation \citep{Derrida86}
described in 
 \citet{ooKauf} book  (p.199 and fig.5.8 
known as `Derrida plot' for $s=2$), expanded to case $s>2$: 
If $a$ denotes a part of system $B$ with the same states of nodes as an 
undisturbed system $A$, then $a^K$ is the probability that the node
has all its $K$ inputs with the same signals in both systems. 
Such nodes will have the same state in the next time point $t+1$. 
The remaining $1-a^K$ part of nodes will have a random state, 
which will be the same as in the second system $A$ with probability $1/s$. 
The part of system which does not differ in $t+1$ is therefore
$a^K+(1-a^K)/s$. It is the same as for RNS \citep{Sole00}. 
The damage $d=1-a$. For $K=2$ we obtain
$d_2= w d_1 - w d_1^2/2 $ where for small $d_1$ we can neglect the
second element.

\subsection{Aggregate of automata - the simplest case of networks for 
coefficient `$w$'}
\label{ch.2.3}

\begin{figure}[ht]
\begin{center}
\includegraphics*[width=7cm]{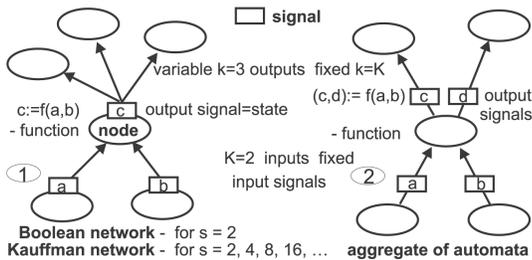}
\end{center}


\caption{The basic elements of Kauffman network (1) and aggregate
of automata  - $aa$ network type (2). Nodes - ovals, signals -
rectangles, links - arrows. Each node transforms incoming (input)
signals into output signals using a function, these signals are
transmitted through links to the next nodes as their input signals. 
$K$ - number of input signals (or links) of a particular node. 
$k$ - number of output links of a particular node (node degree). 
For a particular node of Kauffman network (case 1 on the left) there 
is one output signal (state of node) which is sent by $k$ output links.
$s$ - number of equally probable variants of signal values (in
Boolean network $s=2$, i.e. true and false). In the case (2) of  the
$aa$ network $k$  is fixed  and each output link has its own signal,
possibly different from others.}

\label{fig:2}     
\end{figure}

In the Kauffman networks all $k$ outputs of a node transmit the same
signal - it is the state of node, the value of its function
(fig.2.1). To understand the coefficient $w$ of damage multiplication we
must average by lots of nodes. It is much simpler and more
intuitive (which is important for introducing such a method into
biology) if each output link of a node has its own signal to transmit,
which need not be the same as on other outputs in the same node.
In such a case the function's argument and value are a $K$- and
$k$-dimensional vectors (fig.2.2). Due to function uniformity it is
useful to fix $K=k$. I have introduced such a network in  
 \citep{paris,hof,krab} where I have named it `aggregate of automata' ($aa$). For this
network if $K=2$ then  $d_2=d_1*w-d_1^2*(s-1)^2/(s+1)/s$ which is
obtained similarly as above. Note, that for small $d_1$ we can
neglect element with $d_1^2$. Theoretical curves for aggregate of
automata for case $s=4$ and $K=k=2$ are also included in
fig.1.1-3. 

These figures show that the level of damage equilibrium for aggregate of automata
is much higher than for Kauffman networks. To expect $a_{aa,t+1}$ - the part of nodes in $aa$ network which does not differ in $t+1$ we can use expectation for Kauffman networks shown in previous chapter. Such $a_{Kauff,t+1}$ describes signals on links of $aa$, 
not states of nodes of $aa$ network which contain $K$ signals:\\
 $a_{aa,t+1}=a_{ Kauff,t}^K+(1-a_{ Kauff,t}^K)/s^K$

\section{Network Features Dependency on $s$ from Simulation Using Reversed-Annealed Algorithm}
\label{ch.3}

\subsection{Project formulation and networks type definition}
\label{ch.3.1}

\subsubsection{ Currently investigated network types in RBN range }
\label{ch.3.1.1}

The Kauffman formula gives useful ability to differentiate $k$ within
the network and to investigate different types of networks which
differ in the distribution of node degree $P(k)$ like
Erd\H{o}s-R\'enyi random networks (RBN or better CRBN \citep{Serra04A}), 
on which Kauffman  had worked, 
or  nowadays famous Barab\'asi-Albert scale-free networks.  
This is because the definition of function does not change if $k$
changes. Due to this reason $K$ was fixed and for a directed
network only the $k$ parameter was used \citep{Kauf69,ooKauf} 
as the degree of a node. 
Currently both the parameters of the investigated Kauffman networks - $k$ and $K$ are typically (not in e.g.   \citep{Serra04A}) flexible \citep{Aldana03,Kauf04,Iguchi07}.
Scale-free networks typically turn out to be more adequate to describe
reality \citep{sf99,sf03,Kos04,Kos06,Crucitti2004,Serra04A,Fortunato2005} than old  
Erd\H{o}s-R\'enyi `Random Networks'. 
Most of contemporary authors stay with Boolean networks due to Kauffman's `life on the edge of chaos' approach and consider SFRBN (Scale-Free Random Boolean Network) \citep{Serra04A, Iguchi07} and EFRBN (Exponential-Fluctuation Random Boolean Network) \citep{Iguchi07} together with old classic RBN (CRBN \citep{Serra04A}).

If we change our approach to prefer chaotic networks and use simply omitted $s \geq 2$,  two interesting questions appear: 
1- Are there any significant differences in damage spreading in the chaotic area (which was not interesting up till now) in different network types? 
2- What is the effect of higher $s$ on damage spreading, i.e. is this parameter important? 

\subsubsection{ Network types used in simulation of RSN }
\label{ch.3.1.2}

I have investigated these questions in a simulation using my simplified
algorithm named `reversed-annealed', dedicated for statistical analysis of damage spreading.
The consistency of its result with annealed model expectation makes this
simplification trustworthy. This algorithm will be used later (in the other 
papers) for a simple and intuitive definition of complexity threshold 
(connected with maturation of chaotic features of network \citep{bgec}),
useful at the end of this path to investigation (also using this
algorithm) of structural tendencies in adaptive evolution of
a complex system. These tendencies describe - among other phenomena - some known and
interesting regularities in ontogeny development or human activity which however have not been explained until now.
In this paper we compare damage spreading in five types of
autonomous networks: `$er$' - random \citep{er},
`$sf$' - scale-free ("Barab\'asi-Albert" \citep{sf99,sf03}), `$ss$' -
single-scale \citep{ss} (for $s=2$ known as respectively: RBN / CRBN, SFRBN and EFRBN),
`$aa$' - aggregate of automata and `$ak$' -
a network similar to $aa$ with fixed $K=k$, but using Kauffman
formula where one state of a node is transmitted by all its outputs.

Why don't I use consequently SFRSN, EFRSN, CRSN and so on? There are two reasons for a different choice: The first one is that such names are too long, there are figures where there is only room for one letter to indicate network type, I use there only second letter of my names which are different. First letters of my names indicate a group. The second reason is that the number of considered network types is large (in the next article there will be ten of them) and named differences become another character, not so stable to fix all names for longer time. Therefore level of RSN, RWN, RNS and RBN inside article should be separated. To close problem of naming I include to RSN the family of aggregate of automata together with the Kauffman network family.   

\subsubsection{ Rules of networks growth }
\label{ch.3.1.3}

Construction of the network simulation has two stages: construction 
of the network and damage investigation in the constant network.
Construction of the network depends on the chosen network type. 
Except for the type `$er$' - Random networks, all networks have a rule of growth.
Aggregate of automata `$aa$' and `$ak$'  needs to draw $K$ links
in order to add a new node. These links are broken and their
beginning parts become inputs to the new node and their ending parts
become its outputs (fig.3.1.).

\begin{figure}
\begin{center}
\includegraphics*[width=7.5cm]{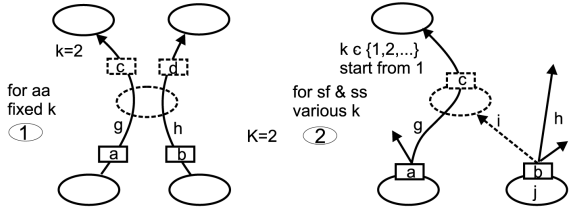}
\end{center}


\caption{Additions patterns for aggregate of automata $aa$ (1) and
Kauffman networks $ss$ and $sf$ (2). ($K=2$) Links g and h (and
function) of node are drawn. Node j is drawn directly instead of
link  h  for  $ss$. For $K>2$ additional inputs are constructed
like the right ones (h or j). The $ak$ network is maintained as
$aa$ but there is only one output signal c (d=c).}

\label{fig:3}     
\end{figure}

For `$ss$' - single-scale network the new node is connected to the
node present in the network with equal probability for each existing node.
For `$sf$' - scale-free network the new node is connected with another
node with probability proportional to its node degree, i.e. to the
$k$ of this existing node. For both types we draw one link first and we
break it like for $aa$ and $ak$ to define one output and its
destination node and the first input. For $sf$ type at least
one such output is necessary to participate in further network
growth. Later we draw the remaining inputs according to the rules
described above - for $ss$ by drawing the node directly, for $sf$ by
drawing a link and using its source node (fig.3.2). 

Damage can take various forms, e.g. in complex computational
networks \citep{nowostawski} but damage spreading in scale free
networks describes typically: epidemic spreading \citep{Kos04}, opinion formation 
 \citep{Fortunato2005,Kos06,Sousa05} and attack or error effects 
 \citep{Crucitti2004,Gallos04}. However, these networks
typically are not directed networks and their important aspect is
the spatial description which uses a particular lattice shape. They
also are constructed in a different way, not only using preferential
attachment \citep{Kos04,Kos06}. 
A partially directed scale-free network was used in   \citep{Stauffer2006} preceded by    \citep{Stauffer2004,Jaco2005}. These networks describe opinion agreement process. In this approach the direction of links is used for construction of a social network and consequently - initiative to contact; however, during opinion exchange an information flows in both directions and each of the talking nodes randomly takes the opinion of its partner. 
This second aspect is more similar to signals flow in Kauffman network, but here it is undirected  and therefore this approach is not similar to Kauffman networks. 
The dynamics of Boolean networks with scale free topology were studied by \citet{Aldana03} and   \citet{Kauf04}, now \citet{Iguchi07}. They look for the difference between the dynamics of $er$ (there called: RBN) and the scale-free random Boolean network (SFRBN). 
 \citet{Iguchi07} used also `directed exponential-fluctuation networks EFRBN' similar to our $ss$.  All they use only $s$=2, flexible $K$ and $k$, therefore their networks differ from ours RSN.

\subsection{Simplified algorithm `reversed-annealed' of damage spreading}
\label{ch.3.2}

\subsubsection{ The main assumption }
\label{ch.3.2.1}

The classic method of damage observation uses two exact processes (quenched) which are compared: $A$ for an unchanged system and $B$ for system with damage initiation \citep{Jan94}. 
I observe one process - only damage spreading, but this process is only statistically correct in a particular range of situations. 
The main assumption is: We consider chaotic system with random function of nodes without memory where damage can fade out only when it is still small, but when it is large, then it grows up to an equilibrium level, where in our algorithm it also stops using the same mechanism which keeps the damage at the equilibrium level (pseudo fade out). In reality (quenched model) it fluctuates around this level infinitely. Our algorithm gives one particular damage size which also fluctuates around this level but for this it needs numerous particular cases of processes and their stops. 
I have found these levels above (ch.2.2, fig.1) using annealed model. They are the validation of presented algorithm. Annealed approximation, however, simplifies the aspect of structure and doesn't consider real fade-out which is important for adaptive evolution. 

\subsubsection{ Limitation }
\label{ch.3.2.2}

The two cases of fade-out described above are mixed, however they have different interpretations. Processes which stop (fade-out) between them with a middle damage $d$ have no interpretation and can be permissible only in negligible frequency. Such cases occur only for $s=2$ (especially for $sf$ 2,3 ($type$ $s,K$) network (fig.5.1) in high level and for $sf$  2,4 and $ss$ 2,3 in small but visible level) which confirms that $s=2$ is extreme. 
The case $s,K = 2,2$ is for every network type out of range of permissible levels of middle damage and we cannot use our algorithm for its investigation - it consists mostly of real fade out cases (fig.1.1) or very low damage equilibrium levels but its 
long tail for higher $d$ is strongly incorrect (too short) in the simulation. 
Too small network size also leads to incorrect results which will be studied in Ref. \citep{bgec}. 
The real fadeout is connected to ordered behaviour and pseudo-fadeout - to chaotic behaviour.

\subsubsection{ Simplifications }
\label{ch.3.2.3}

In this algorithm, we only calculate nodes with one or more changed input signals. 
To detect changed input signals, signals are first calculated and memorized during network growth. 
If we detect a changed input signal as the effect of damage initiation, we do not care what remaining input signals are. They can be changed before or after the calculation of this particular node (in damage calculation),  e.g. as effect of feedbacks loop. If a node is affected by damage, which means that at least one input signal was changed, then the node function is calculated (if necessary but not in situations described in this paper) using `old' remaining input signals, but only once (i.e. a node is never calculated twice). 

The damaged part will become a tree.
In this paper we also will not use concrete functions for nodes. If the input state is changed, then the output state is random.  
This calculation gives an answer, whether output signals of this node have undergone any changes. If its input signals change later, then it will not be calculated next time - for statistically correct damaged area it is not needed. Any initiation of a particular node in a particular network should lead to statistically the same damaged area but in each particular case it may be different. 
We, however, are not interested in a particular case but in a statistical result.
Such an algorithm works fast and gives correct statistical effects. 

\subsubsection{ Intuition behind }
\label{ch.3.2.4}

An intuition behind this algorithm can be found when we consider a network 
without feedbacks, where each signal on the node output is equal to the 
value of the function of current signals on the node inputs. It is not 
a typical system state - in the next time step (e.g. in the synchronous mode)
nothing will be changed. 
Now we introduce a disturbance changing one node function (permanently) and we calculate this node. To obtain a new stable state of the system we must calculate only nodes whose at least one input signal had changed (as a result of damage). For this calculation we can use the old signals on the remaining inputs if for a given node they do not depend on the remaining nodes waiting for calculation.
Such a node will always exist because a node does not depend on itself. After a finite number of node calculations all the node states will be equal to the function value of current node inputs as was the case at the beginning, however lots of node states will be different than at the beginning, i.e. - damaged. As was mentioned above we do not use concrete functions, therefore we need not to check a dependency and we can calculate any waiting node.

In the case with feedbacks sometimes an already calculated node gets a damaged input signal for a second time. For measuring the statistical effect only it is not necessary to examine its initiation for the second time, however, if such second initiation will be processed, then the process may never stop.

\subsubsection{ Experiment structure }
\label{ch.3.2.5}

In this paper we investigate the damage in a system of a particular size.
When a network achieves the assumed number $N$ of nodes we stop the growth and we start to initiate damage:  we change the output state of each node using all remaining possibilities as damage initiation. It is the smallest initiation and in the first few steps the damage can fade out. 
It is a real fade out of damage. In this short way damage can meet an already damaged node which is not calculated for the second time (which helps damage to fade out), however, such an event has a very small probability. We assume that if damage fades out when it is small, then it is not due to meeting an already calculated node. This is a simplification of our algorithm. In this case the number of damaged nodes is interpreted as the number of damaged nodes during the whole process from initiation to real fade out.
If coefficient $w>1$ then on average the damage grows. If damage is great, i.e. the number of nodes with changed output state is large (the number of calculated nodes due to their input state was changed is also large), then it practically cannot fade out (probability of such events is very low, we neglect them), but during this damage growth there are less and less nodes which are not reached by damage yet. Therefore the avalanche of damage must slow down and stop (the growth). It looks like a fadeout, but it is equivalent to the achievement of the stable level by the damage which appears at the end of curves in fig.1.2 or on cross of curves with line `$d_{t+1}=d_t$' in fig.1.1. This level is an equilibrium state, as fig.1.1 shows. In our simplification the process stops at this level due to the `pseudo-fade out' on already damaged nodes. 
Now the number of damaged nodes is interpreted as the 
equilibrium level 
and it describes the statistical state of the system at any one   time step 
after the equilibrium is reached.

The number of nodes with changed output state (i.e. number of damaged nodes) divided by $N$ is equivalent to the damage size $d$, despite the fact that they are damaged during the whole process (using our algorithm), not only in the last time step. Note, if damage fades out, regardless of the way it happens (pseudo 
or real fade-out), then in the last time step  
damage size looks like zero.  
However, such a view is incorrect 
in the case of pseudo fade out, it does not take into consideration the fact that we do not recalculate a damaged node when its input signal changes for a second time. Such a false suggestion appears due to the simplification of our algorithm. 

\subsubsection{ Reversed-annealed approximation }
\label{ch.3.2.6}

We calculate the damage using a fi-fo queue (first in - first out) for nodes with changed input signals waiting for calculation. The queue length in time $t$ (as in the synchronous mode) dependency is very similar to the one shown in fig.1.3. Small differences at the point of maximum are shown in table 1. The time step number $t$ is defined observing this queue but for control of the process it is not necessary.
Simplification used for our algorithm is a step from the `quenched model'  \citep{ooKauf} in the direction similar to `annealed model'  \citep{Derrida86,Derrida86W}, however it is a small deviation only and in `reversed' direction. These simplifications are: 1 - only nodes with damaged input signal are calculated; \\
2 - second initiation of already calculated node is neglected; 3 - when function is not used, then new node state is random, else for the remaining, not damaged input signals old values are used.
A `reversed annealed simulation' was performed: We randomly define new state for all nodes keeping old network structure and we obtain very similar results. 
Because the input sites remain constant over time they are quenched \citep{Derrida86W}, however, the functions are simplified therefore the algorithm is not quenched but (reversed-) annealed. The dynamics of such a net is thus approximated in a non-deterministic way.
Note, the task which we define for this algorithm is not an adaptation or optimisation of network  function but the investigation of damage spreading (in this paper) and (in next papers \citep{bgec}) - of statistical properties of changes leading to smaller or larger damage which concerns changes constituting adaptive evolution. It will look there like a greedy algorithm, however, `greedy' property is not well adequate for its description. Our algorithm is much simpler than the quenched one, therefore it is computed much faster. By yielding one particular result on external outputs (instead of a circular attractor) it allows to define fitness in a simple way and so this algorithm opens new areas of investigation (e.g. structural tendencies).

\subsection{Simulation effects and comparison to theoretical
expectations}
\label{ch.3.3}

\subsubsection{ Simulation parameters }
\label{ch.3.3.1}

The computer program realizing the above algorithm is prepared for $s=$ 2, 3, 4, 8, 16, 32, 64 and $K=$ 2, 3, 4. 
I have investigated the whole of this area but the most interesting part is the area near the phase transition from chaos to order where differences are larger. For comparison of dependency in main parameters $s$ and $K$ I choose five cases described as $s,K$: 2,3; 2,4; 3,2; 4,2; 4,3 for five
network types described above: $er, ss, sf, ak, aa$. 
In this set there are: $K=3$ and $4$ for $s=2$, next: $s=3$ and $4$ for $K=2$. Similarly for $K=3$ and $s=4$ the second parameter has two variants. Cases 2,3 and 4,2 have the same $w=1.5$. The coefficient $w$ is the smallest for case 3,2 (1.33) and the largest in the shown set for 4,3 (2.25). The simulation results are shown in fig.4, fig.5 and fig.6, also in table 1 but only in the main fig.6 the whole set of the above enumerated cases is shown.
For the most interesting networks $sf$ and $er$ the full set of combination $s$ and $K$ in range of values 2, 3, 4 is show in fig.6.3.
Each simulation consists of 600 000 damage initiations - e.g. for $s=4$
(excluding $aa$) 100 different networks grow randomly up to
$N=2000$ (or $N=3000$) nodes and later each node has its output
state changed 3 times (2 times). For $aa$ and $s=4$ I use 20 networks and the output state is changed 15 times.

\subsubsection{ Comparison results to expectations }
\label{ch.3.3.2}

\begin{table}[b]
\centering \caption{Some simulation results and their comparison
to the theoretical (th.) expectations showed in fig.1. (For 4,2 d1 
shown  for $aa$ and remaining) $s,K=$ 2,3 4,2 4,3 d1 - $dmx$
(equilibrium level of damage size, see fig.1) taken from maximum
position of the right peak in $P(d)$ distribution where $d$ is the
damage size when damage fades out (in the sense of our algorithm)
d2 - $dmx$ taken from the first stable maximum (plateau) of $d$ in
$d(t)$ (fig.1.2 and fig.1.4) t1 - position of maximum in
increasing $d(t)$ (fig.1.3). Typically lower in simulation. t2 -
visible range of the right peak in increasing $d(t)$ (fig.1.3).
Typically higher in simulation. t3 - position of maximum of the
right peak in $P(t)$ distribution where $t$ is the time step number
when damage fades out (in the sense of our algorithm)}


\begin{tabular}{rrrrrrrrrrrrrrrr}
\noalign{\smallskip}\hline\noalign{\smallskip} 
net &2,3 d1 & d2 & t1 & t2 & t3&4,2 d1 & d2 & t1 & t2 & t3&4,3 d1 & d2 & t1 & t2 & t3 \\

\noalign{\smallskip}\hline\noalign{\smallskip}

 $aa$ & 1338  &  1319  &  16 & 35 & 28  & 1668 &  1672 &   17 & 32 & 27&  1938 &1908  &  10 & 16 & 15\\
 $ak$ & 772 & 690 & 14 & 34 & 25  & 1335 &1307 & 16 & 31 &26& 1473 & 1515 & 10 & 15 & 14\\
 $er$ & 770 & 880 & 14 & 34 & 25 &  1335 & 1336 & 15 &  34 &  24& 1473 & 1441 & 10 & 17 & 14\\
 $ss$ & 774 & 654 & 13 & 40 & 24 &  1335 & 1202 &    16 &  36 & 25 & 1476 & 1495 & 10 & 19 & 14\\
 $sf$ & 818 & 877 & 12 & 45 & 21   &   1344 &   1217 &    16 & 43 & 26 &  1476 &  1485 & 10 & 26 & 15\\
th.  & 764 & & 17 & 26  & &$aa$1667 & 1333 & 18 & 28& &1472 & & 10 & 14\\

\noalign{\smallskip}\hline
\end{tabular}
\end{table}

As the first effect of simulation I show a comparison  
to the expectation presented in fig.1.3 and fig.1.2. The obtained
distributions of increase of damage in consecutive time steps in
synchronous mode of calculation are very similar to the ones shown in
fig.1.3. (A small difference at the point of maximum is  shown in table 1).  
However this similarity conceals some statistical diversity of
speed of damage spreading process. This diversity appears when we try to compare 
obtained results especially for networks $sf$ and $ss$ (e.g. fig.1.4) to fig.1.2.  
In fig.1.4 a case for $ss$ 2,3 is presented and a similar case $sf$
4,2; they are both less rough than $sf$ 2,3 and $sf$ 3,2. We can identify a few
independent processes with different speeds of damage growth. 
This diagram is plotted for each particular case of damage initiation replacing the old one.
When damage reaches the level of equilibrium it stops (in our simplified algorithm) 
and there is no data for later time steps. In this later area, therefore, we can observe
other processes which are earlier and slower. For network types
scale-free and single-scale this speed is strongly connected with
the time of reaching the hubs by damage. For network types $er$ and
obviously $ak$ and $aa$ there are no hubs and the obtained curves are
more uniform and similar to the theoretical ones shown in fig.1.2.

\subsubsection{ Distribution $P(t)$ of time of damage fades out }
\label{ch.3.3.3}

\begin{figure*}
\begin{center}
\includegraphics*[width=16cm]{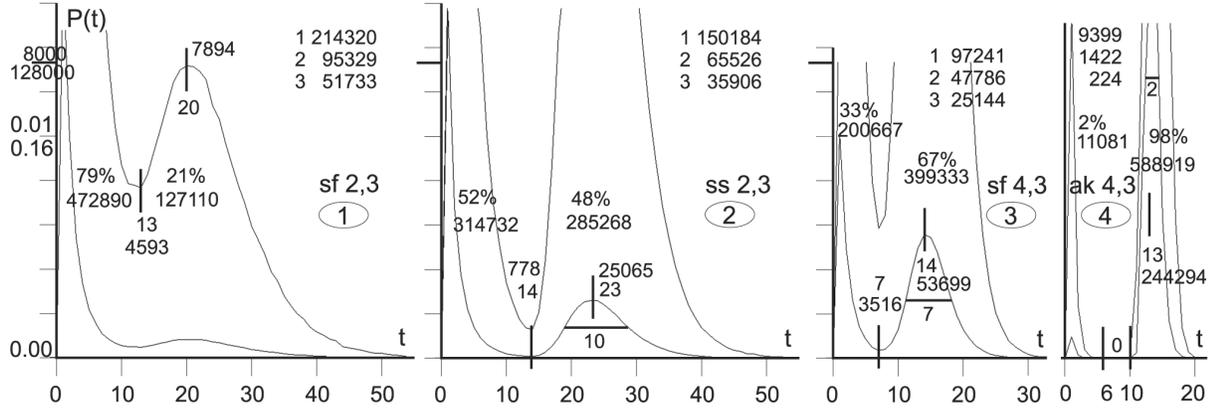}  
\end{center}


\caption{Distribution $P(t)$ of time $t$ when damage fades out 
(in the sense of our algorithm). 
All scales are the same but there are two scales for probability $P$.
Figures are ordered from low to high level of chaos but there are 
3 variable parameters: network type, $s$ and $K$. Here for all cases $K=3$.
In the first two (1, 2) figures $s=2$ but the network type differs - $ss$ is less ordered 
(left peak =52\%) than $sf$ (79\%) (see fig.6.2).
In the last two (3, 4) figures $s=4$. Fig.(3) shows $sf$ network which is the less 
chaotic one. Fig.(4) shows $ak$ network which is the most chaotic one (in Kauffman 
mode). Here for the first time there appears a gap of zero frequency between peaks.
The remaining $ss$ and $er$ cases are between them (see fig.6). 
All the networks contain 2000 nodes. 
All the distributions are obtained from 600000 events of damage initiation.  
In the numbers there are shown: the positions and values of minima
between peaks and the right maximum; the width of 
the right peaks at half of their height; a few
of the first values for the left peaks; the number of events in both peaks
and the percent of all the 600000 events in each peak (this important 
information is hard to estimate only from the shown curves). }

\label{fig:4}     
\end{figure*}

Fig.4 shows the distribution of time of damage
fadeout in both real and `pseudo' cases. There are two peaks in
this distribution: one for early fadeout in the first steps 
(we interpret them as real fadeout) and the second one later, 
for `pseudo-fadeout' when damage reaches the 
equilibrium level which is high (at the last time steps in our algorithm). 
For the network cases with wide
range of node degrees like $sf$ and $ss$ with a large fraction of
$k=1$ nodes the probability of early fadeout is much greater especially
for small $s=2$. If $K=3$ then 60\% nodes for $sf$  and  33\% for
$ss$ have $k=1$ but there are 11\% and 20\% nodes of $k>4$ which
have 55\% and 46\% outgoing links respectively. 
If $K=2$ then respectively for $sf$ and $ss$ networks, 
67\% nodes and 50\% nodes have $k=1$, there are 7\% and 6\% nodes of
$k>4$ which have 34\% ($sf$) and 19\% ($ss$)outgoing links . For $s=2$ nodes
with $k=1$ have $w=1/2$ but for $s=4$ there is $w=3/4$. 
To reach the threshold value 1 for $s=2$  there remains two times
 more (1/2) than for $s=4$ (1/4).
Only nodes with higher $k$ can increase $w$ 
and reach this threshold but they are rare as shown above.
Therefore locally it seems to be $w<1$ (like for the ordered regime) more often 
for $s=2$  and early fadeout is more probable then than for $s=4$. 
Hubs are present in this case. The biggest hub ($k=955$)
appears in $sf$ when $K=4$, for $K=3$ it reaches  $k=520$. 
This single hub takes 12\% (the second 9\%) of all the outgoing links. 
Hubs decrease the average $k$ and in effect also the average $w$ for 
remaining nodes, which makes the remaining area `ordered' and helps damage 
to fade out before the first hub is achieved.
For $er$ network even $k=0$ occurred but nodes with $k<2$
constitute less than 1/4 of all the nodes. If $s$ is small, e.g.
$s=2$, then the coefficient $w$ is locally especially low. Note
that we have used local coefficient $w$ for the explanation. On the
opposite end (only of Kauffman mode, i.e. excluding $aa$ network) the case 
of $ak$ 4,3  lies where $k<2$ and hubs are absent. In this case the coefficient 
$w$ of damage propagation is high and equal for all the nodes, 
therefore the early fadeout is very small and most of the damage 
grows until the equilibrium level is reached.

Fig.4 consist of four different distributions.
They are all plotted for networks of $N=2000$ nodes, but this sequence
looks like a sequence of distributions of stage of e.g. $ak$ 4,3
growth. (We investigate this process in Ref.\citep{bgec} in much more detail. 
It can be a base for the definition of complexity threshold.) 
This means that $sf$ 2,3 in left (fig.4.1) looks like small, not yet mature networks
$ak$ or $er$. In an $sf$ network 2,3 in the most cases (80\%) 
the damage fades out (real fade out) 
without reaching hubs. Such damage behaviour is more typical
of `ordered' networks rather than `chaotic' ones. 
This is `liquid region' available for adaptive evolution. 
For $ss$ 2,3 and $sf$ 4,2 both the peaks have a similar area (fig.4.2) but the left
peak is much higher. Later, (fig.4.3) for $sf$ 4,3 and (fig.4.4) $ak$ 4,3
the left peak has a smaller area and the right
peak which describes chaotic behaviour contains most of the cases.
This looks like advanced stages of transition from order to chaos. 
In the percolation theory maximum of right peak describe size of large cluster.
It appears much after percolation point in which only left peak exists 
and it just reached power-law shape.

\subsubsection{ Distribution $P(d)$ of fade out in `damage size'}
\label{ch.3.3.4}

\begin{figure*}
\begin{center}
\includegraphics*[width=16cm]{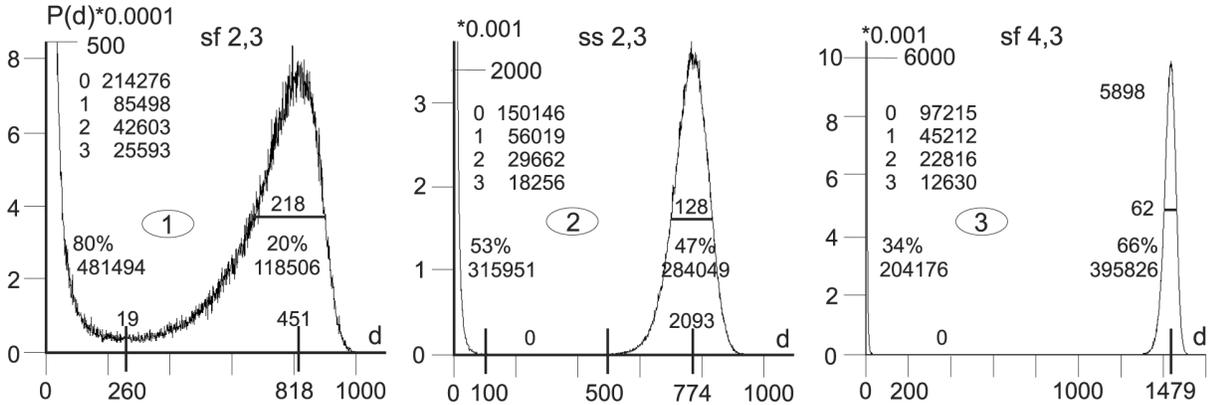}
\end{center}


\caption{Distribution $P(d)$ of `damage size'  when damage really 
or `pseudo' (via stabilization at the equilibrium level) fades out. 
Definitions and interpretations in the text. Parameters of experiment and the shown 
cases (without last) as in fig.4. Scales are different but shape is interesting.
Numbers show the same values as in fig.4. 
Except the extreme case (1) $sf 2,3$,  the peaks are clearly separated 
which allows to measure the levels of order and chaos more clearly.
E.g. in (3) a left peak exists, it contains 34\% of the
events but it is hard to see.}

\label{fig:5}     
\end{figure*}

The phenomenon of different speed of damage spreading described
above is the reason behind the large width of these peaks and lack of sharp boundary
between them here (fig.4.1-3).  This remark suggests that the variable
$t$ - time of  damage fadeout is not the best choice. However,
the variable $t$ is interesting in practice and therefore often used
 \citep{Jan94}. A similar distribution of damage fadeout in the variable:
damage size $d$, shown in fig.5.1-3 appears much
better suited for the description and understanding of the underlying
mechanisms. 
\citet{Serra04} have introduced term `avalanche' which is consequently used later \citep{Shmul05,Ramo06,Serra10},  for damage size measured in damaged node number. Such a variable is used in fig.5, but above used parameter $d$ differ from `avalanche' only in normalisation ($d=avalanche/N$) and we will stay with $d$ only.
Using such a variable we also obtain the same two peaks
but this time they are very narrow and a big segment of
exact zero frequency lies between them typically. 
The only exception from this rule is the extreme case of $sf$ 2,3 (fig.5.1) 
but we have discussed the causes of this exception above. The case of 
$sf$ 3,2 and (fig.5.2.) $ss$ 2,3 follows the rule  but the second peak is not
very narrow and we can find some single cases between the peaks.
(These cases have no clear interpretation in our algorithm.)
All the remaining cases are similar to the last $sf$ 4,3 shown in
fig.5.3, small differences concern proportion of both peaks and
the peaks' width.

Position of maximum of the second peak is exactly equal to the
theoretical point of equilibrium of damage size ($dmx$), it is obviously 
much more exact than the values of maximum which can be read from the distribution
like in fig.1.4 which are for one particular process. 
The comparison of this 3 values for different cases is shown in table 1.

\subsubsection{ $w$ doesn't substitute consideration of $s$ and $K$ separately }
\label{ch.3.3.5}

\begin{figure*}
\begin{center}
\includegraphics*[width=16cm]{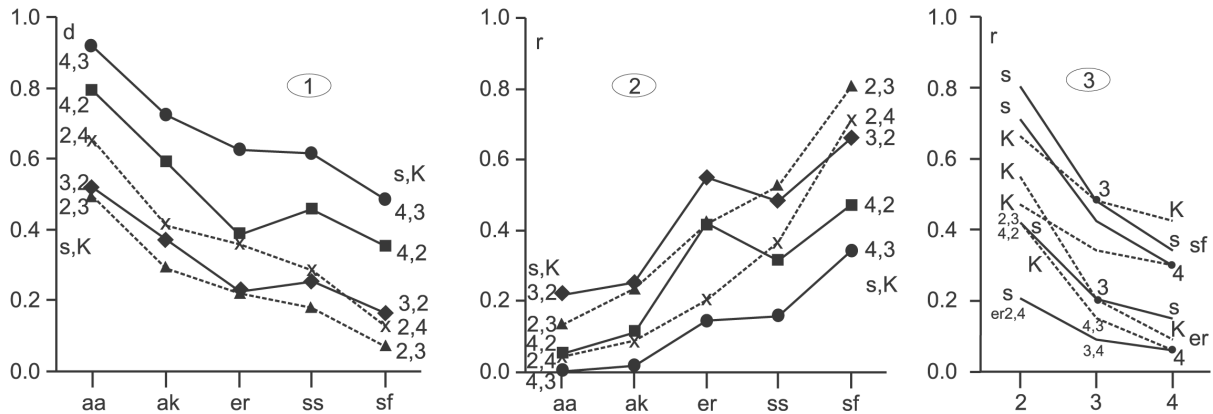}
\end{center}


\caption{Average (for all initiations) damage size $d$ at the process end (1) 
and real (early) fadeout as part $r$ of all initiated processes - `degree of order' (2) 
for five different network types and small values of parameters $s$ and $K$. 
Comparison of slopes of dependencies $r$ on $s$ and $K$ for $sf$ and $er$ networks (3). 
The points have 3 decimal digits of precision. Note, $d$ considers also early (real)
fadeout (for which $d=0$ is taken), not only the equilibrium level for large damage.
Cases of parameters $s$ and $K$ are selected for easy comparison of dependency on them.}

\label{fig:6}     
\end{figure*}

As it was discussed above, the new network types, especially the
scale-free networks, due to concentration  of a great part of links in
a few hubs, exhibit significant differences in behaviour of damage
spreading. These differences appear especially near the boundary of
chaos and order and are more intensive for $s=2$. To summarize these
differences I show fig.6.1 where we can compare average (for all initiations) 
damage size $d$ for each simulated case of network type and $s,K$. 

Depicted data have 3 decimal digits of precision, therefore the shown
differences are not statistical fluctuations. As it can be seen,
using higher $s=4$ for $K=2$ causes different behaviour of damage
spreading than for $s=2$ and $K=3$, especially for $er$ network type, 
despite the 
same value of coefficient $w=1.5$, therefore these both parameters cannot
substitute each other, i.e. we cannot limit ourselves to one of
them or to the coefficient $w$. 

\subsubsection{ `Degree of order' of network }
\label{ch.3.3.6}

Fig.6.1 contains two different causes which differ results. 
One of them was already described at the end of ch.2.2. 
It is the different level of damage equilibrium visible in the
theoretically obtained  fig.1. If we remove this aspect from fig.6.1, then 
what remains is the second cause which is connected to the early (real) 
fadeout. This cause depends mainly on the network type and the $s$ parameter 
which is depicted in fig.6.2. This figure shows how big is the part of 
initiations which end in real fadeout. 
They are separated using threshold on $d=250/N$.
Compare the point for $sf$ 2,3 to the fig.4.1.
This aspect contains the mechanism of $er$ distinctness which results from the 
events $k=0$. 

Fig.6.2 has an important interpretation: the depicted parameter $r$ is a `degree of order' of a network in a certain aspect. It describes the probability of real fadeout of damage. However, for application to living object description it should be noted, that this fadeout only occurs in a random way which does not consider negative feedbacks collected by adaptive evolution. 

These investigations using simulations of different network types were designed 
and included in this paper to show that the parameter $s$ is important and we cannot 
limit ourselves to the parameter $K$ only when we study chaotic behaviour.
The dependency on $s$ is about as strong as the dependency on $K$ but it also differs 
from dependency on $K$ for different network types. 
In the aggregate of automata the state of a node has $s^K$ variants and this 
network type has obviously stronger (and different) dependency on these parameters 
than Kauffman networks. 
The $ss$ and $ak$ networks exhibit symmetrical dependency in $s$ and $K$ 
but for the most interesting $sf$ and $er$ network types there is no symmetry, 
which is depicted in the fig.6.3. 
For $sf$ the dependency on $s$ is stronger but for $er$ - weaker than the dependency 
on $K$.
For each network type two sets of two `lines' are shown. 
Each `line' consists of three cases for the same value of parameter $s$ or $K$ and all three values (2, 3, 4) of the second parameter. 
E.g. we compare the line consisting of (s,K) cases 3,2 3,3 3,4 and the line of 2,3 3,3 4,3 cases. 
These two lines have one common case 3,3 indicated in the figure. 
We are interested in the slopes of both compared lines. 
The relative slopes appear to be approximately constant for both parts of each lines. 
For both network types shown  the slopes for parameters $s$ and $K$ are significantly different
which already for two steps leads to significantly different values. 
These differences are not big but may be important.
The significantly lower damage size for $sf$ network which can be seen 
in fig. 5 is known 
 \citep{Gallos04,Crucitti2004} as the higher tolerance of a scale-free 
network of attack. 
Also 
 \citet{Iguchi07} state: ``It is important to note that the SFRBN is more ordered than the RBN compared with the cases with $K = \langle k\rangle$'' which in our fig.6.2 is clearly visible, however in the area shifted into chaotic direction.

\section{Conclusion}
\label{ch.4}
There are three different types of conclusion of this paper. 

The first is mainly interpretative - systems describing living objects should be chaotic, not ordered.
More exactly - they should be much more chaotic, even purely (maturely) chaotic (with possibility of damage avalanche which models death) than Kauffman expects in his ``bold hypothesis: Living systems exist in solid regime near the edge of chaos''. There, living objects are on opposite shore of Kauffman's `liquid region near edge of chaos'. In this region,  a part of small changes, available for adaptive evolution, exists, which I show as `degree of order' in simulation results. 
Death is possible in all living objects and is necessary for Darwinian elimination. It should be an element of the model. Damage avalanche which can only model death is available only in chaotic phase. The stability of gene regulatory network is the main experimental argument behind Kauffman's localization of living systems on order-chaos axis but in this model negative feedbacks are not considered sufficiently.

The second, main conclusion of this paper is that the parameter $s$ (number of 
equally probable variants of signal i.e. states of nodes in Kauffman network) 
is an important one. It approximates other statistical mechanisms 
than using bias $p$ for two variants only. 
This new mechanism is named here Random Signal Network (RSN). 
(Full name: `Random equally probable Signal variants Network' is too long.)
Together with RNS and RWN it expands the notion of `Kauffman network' 
which was the synonym of `Boolean network'.
For the investigation of damage spreading behaviour in chaotic phase, 
especially for scale-free networks, 
parameter $s$ cannot be neglected or substituted by the parameter 
$K$ (number of node inputs) and others like $P$ 
(internal homogeneity) and bias $p$ (the probability of one of two signals).
It ($s>2$)  leads to much higher levels of damage equilibrium which can be 
found theoretically using extended annealed approximation. 
It also leads to different `degree of order' for different network types 
which is seen in simulation.
Next arguments of various types are collect that especially for adaptive 
systems describing real living or human-designed systems, 
the $s$ parameter should generally be greater than two 
which placed these systems (with constant $K>1$) in the chaotic area. 
I believe that the typically observed case of two alternatives with different probability
has typically bases whose description using bias $p$ is incorrect 
and $s$ should be used instead.

Along the way of investigation of properties of the parameter $s$ using
simulation, I have found that the contemporary first BA 
scale-free network has significantly different behaviour in damage
spreading than the Erd\H{o}s-R\'enyi random network which was used 
in Kauffman's path and these differences are largest for $s=2$.
Generally, the networks with higher frequency of $k<2$ nodes ($k$ is node
degree - the number of node outputs, for autonomous systems $K=\langle k\rangle$ ) have 
higher chances of damage fade out in the critical
beginning period. If hubs are present then this chance also
increases because they decrease the average $k$ for remaining nodes
which helps the damage to fade out before the first hub is reached.

For this simulation I design and describe here a special simplified algorithm 
which I name `reversed annealed'. I also use it for the next investigation, 
of `complexity threshold' and `structural tendencies'. 
This algorithm uses only calculation of damage spreading up to 
reaching the equilibrium level instead of calculation and 
comparison of two systems - damaged and undisturbed one as in the classic method.
Therefore it is much more optimal algorithm. 
And now the third conclusion of this article - the technical one.  
This simplification has small meaning for investigations presented in this paper, 
however, it is a critical key to the area of `structural tendencies' in adaptive evolution 
which are the aim of the path starting with this paper (see also \citep{bgec}).
The main feature of this algorithm is that it gives a single particular output state, 
however only statistically correct, instead of long cyclic attractors. 
In my approach avoiding attractors is useful in opposition to Kauffman's approach 
where properties of attractors are very interesting.

Another technical useful result of this paper is
the coefficient of damage propagation, which I introduce here. 
It connects two main parameters $K$ and $s$ and describes the first, 
critical period of damage spreading (in the first theoretical approximation). 
It is simple and intuitive and it easily shows when damage should explode: 
damage should explode always if $k>2$ or $s>2$. 
The ability to produce an explosion of damage is one of the definitions 
of a chaos which I used following Kauffman for chaotic systems. 
The above mentioned influence of the hubs suggests that such a coefficient 
should be considered more locally, e.g. in the area of damage spreading.

\section*{Acknowledgments}
Research financed by Polish government as grant 3 T11F 035 30.

\section*{Appendix A - Thermostat}
\label{ch.Appendix A}

\begin{figure}[b]
\begin{center}
\includegraphics*[width=8cm]{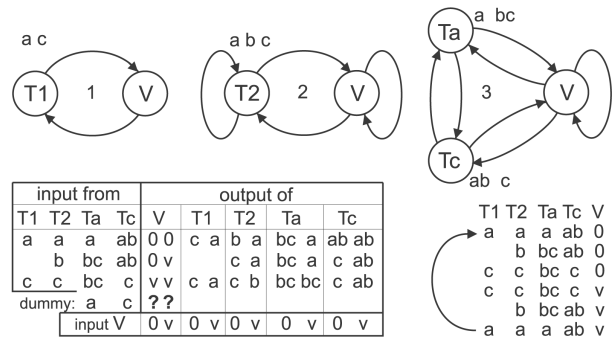}
\end{center}


\caption{Thermostat of fridge described using Kauffman networks as example 
of regulation based on negative feedbacks. 
Case (2) describes it just as it is in reality - temperature $T2$ is split into three sections $a$ - too cold, $b$ - accurate, $c$ - too hot but therefore this case is not Boolean.
To hold signal in Boolean range we can neglect temperature state $b$ - case (1) or split node $T2$ into two nodes with separate states - case (3) which together describe all temperature states but this way dummy variant ($a,c$) of temperature state is introduced. 
Node $V$ decides power for aggregate: v - on, 0 - off. 
Common tables of functions for nodes (left one) and for consecutive state (right one) are attached. In left table for case (1) and (2) certain lines of $V$ states should be omitted. Similarly in right table for case (1) there are only four lines.}

\label{fig:7}     
\end{figure}

We try to check how Kauffman's assumptions for regulation based on negative feedbacks work. 
Let us analyze a simple known example of thermostat in fridge.
If fridge leaves proper temperature of range $b$ as result of environment influence and enters too high temperature of range $c$, then power supply for aggregate is turned on and temperature inside fridge goes down. It passes range $b$ and reaches too cold range $a$, then power is made off and temperature slowly grows through $b$ section. Case (2) in fig.7 properly describes this regulation mechanism in Kauffman network terms. However, there are three states of temperature $a,b,c$ which are described by three variants of state of node $T2$ and therefore this case of Kauffman network is not a Boolean network.

To hold signal describing temperature in Boolean range (two variants only) we can neglect temperature state $b$. This is case (1) but here, the most important proper temperature state, where fridge stays most of time, is missing. Almost any time we check the state of a real fridge this state is not present in such description. Reading such description we find that wrong temperature $a$ - too cold occurs directly after wrong temperature $c$ - too hot and similarly in opposite direction. Is this fully correct information and description?

Splitting node $T2$ into two nodes $Ta$ and $Tc$ with separate states - case (3) is the second method to hold Boolean signals. Two separate Boolean signals create together four variants but temperature takes only three. New dummy state emerges: $a$ - too cold and $c$ -too hot simultaneously. It has no sense and never appears in reality but function should be defined for such a state. In table shown in fig.7 the functions values for nodes of case (3) for such dummy input state are marked by `?'. Anything can be put there, however, for statistical investigation it is taken as real proper state.

Do  cases (1) and (3) describe reality adequately? I'm afraid that the answer is `NO'. It is because Boolean networks are not generally adequate. We can describe all what we need using Boolean networks but in lot of cases we will introduce dummy states or we will simplify something which we don't like to simplify. In both cases statistical investigation will be not correctly connected to reality. The only way is to use real number of signal variants and not to limit ourselves to only two Boolean alternatives.

Even in simplest case (1), but also in remaining cases, temperature as `essential variable' varies and this is necessary for regulation. For feedback mechanism it cannot be fixed. See right table in fig.7 which shows consecutive states for nodes in all cases. If we fix temperature constant e.g. in range $b$ which is a good approximation measuring it frequency in time, then it can be treated as effect of regulation mechanism work but it deprives of job existing in network feedbacks. Such mechanisms keeping it constant are there outside such a network. In effect of fixing `essential variables' values Kauffman removed stability mechanisms of these variables from such networks and cannot look for cause of this stability inside investigated networks. Found `spontaneous order' and `homeostatic stability' of ordered phase is not this searched cause.

\section*{Appendix B - main Kauffman's terms }
\label{ch.Appendix B}
 
This paper discuss certain details in Kauffman's approach and can be comprehensible only for readers familiar with this area, however, for completeness I introduce some short remaining of main terms which may be understood a little bit differently.

I use the term `chaos' as 
 \citet{ooKauf} does. 
It differs from the more common definition \citep{Schuster84} 
used for continuous arguments of a function. 
In networks there is a finite number $N$ of nodes, each one with a small natural number $s$ of possible states. Such chaotic systems differ from the ordered ones in damage spreading.

Damage (or `perturbations' \citep{Serra04,Ramo06}) is the difference between two identical functioning (dynamical) systems which appears as an effect of some disturbance  in one of these systems \citep{Jan94}. Typically a very small change is investigated which initiates damage, e.g. a change of state of one element of the system. For chaotic systems a small initiation of damage typically causes a large avalanche of damage which spreads onto a big part of the system, however, it ends at an equilibrium level. 
The existence of this equilibrium level as the limit of damage growth is the main difference between this `chaos' and the more commonly used definition. Levels of damage equilibrium are discussed in ch.2.2 basing on annealed approximation \citep{Derrida86, ooKauf}.
High stability of ordered systems does not allow for damage to evolve into a large avalanche. There could be very small avalanche  in `small unfrozen lakes of activity' which I don't name `avalanche'
(however \citet{Serra04} do), especially - `large avalanche'.

There are four similar terms used in the Kauffman approach which should be clearly separated: `structural stability', `homeostatic stability', `ultrastability' and `spontaneous order'.

As it is described in Ref.\citep{ooKauf}, the `structurally stable systems' evolve in correlated landscape (e.g. of fitness) which typically allows the small initial change to give a small change as its effect. This landscape is considered on a space of system parameters where the nearest neighbours vary by the smallest change of connections, function or states. It means that such neighbours are similar and they function similarly.
Chaotic systems, which are not `structurally stable', evolve in uncorrelated landscape where small change typically causes crossing many walls of bifurcation, radically changing the system's properties (e.g. adding a new basin of attraction). Function of neighbours typically are not similar.
Adaptive evolution needs small changes. Large change typically lost collected aptness. Kauffman approach looks therefore for such `structurally stable' area and found that it occurs between chaotic and ordered phase. 
It is the cause why in the Kauffman's approach the phase transition between chaos and order is one of the most important themes of investigation. 
Kauffman isolates this area basing on implication for evolvability and structure of fitness landscape as third region named `liquid' between `solid' (because frozen - ch.1.2.2) ordered region and `gas' - chaotic one. ``Landscape is very rugged in the chaotic regime. This ruggedness is a direct consequence of the fact that damage spreads widely in networks in the chaotic regime. Almost any single mutation will dramatically alter landscape structure'' \citep{ooKauf}. Note - not all mutation. Liquid region lies on formally ordered and chaotic regions and its boundaries are smooth and not well defined. \citep{Holland98} named this region `the edge-of-chaos membrane'.

When initiation of damage occurs then effected damage (change) can be large or small. Minimizing of initiation's effect is the homeostatic feature. Typical homeostatic mechanisms based on negative feedbacks are named `ultrastability' in Kauffman approach and are considered separately (see ch.1.2.2). Term `homeostatic stability' Kauffman uses for general resistance of system to disturbance. This term is not formally introduced in   \citep{ooKauf}, it is absent in index there, however, it is used mainly for stability which emerges spontaneously together with spontaneous order which are main topic of Kauffman researches. Due to practical rejection of ultrastability from this researches (ch.1.2.2 and Appendix A) homeostatic stability contain there only aspect of spontaneous  resistance to disturbance.

The ordered area, where maximum of spontaneous order (ch.1.2.1) occurs, exhibits also highest `structural stability' which is useful for adaptive evolution and `homeostatic stability' which should be an effect of evolution. 
The maximum of spontaneous order is one of the most important features of this area, especially 
when spontaneous order is taken to be the real one (ch.1.2.1).
For Boolean networks it is the case of $K=2$:  ``If the stability of each state cycle attractor is probed by transient reversing of the activity of each element in each state of the state cycle, then 80-90\% of all such perturbations, the system flows back to the same state cycle. Thus state cycles are inherently stable to most minimal transient perturbation'' \citep{ooKauf}. 

Evolution using random walk in the space of network parameters should tend to this area because there effects of evolution are larger. This expectation is known as `life on the edge of chaos'. 
Kauffman even expects of living systems in solid regime in his ``bold hypothesis: Living systems exist in solid regime near the edge of chaos, and natural selection achieves and sustains such a poised state'' - page 232 in   \citep{ooKauf}.

A large change - large damage avalanche in chaotic systems is obviously taken as improbable in adaptive evolution but conclusion that adaptive evolution is improbable in chaotic systems would be too quick. As noted above (in description of liquid region) in chaotic systems small change only typically causes large effective change. There can also happen small effective changes which may be used by adaptive evolution. It happens more frequently if parameters of network are nearer phase transition between order and chaos. This possibility is discussed in ch.3.3 in fig.6.2 as degree of order of different networks and is one of more important theme of this paper.

In fig.8 emergence of chaos during network ($er$ 4,2 see ch.3.1.2. \& ch.3.2.2.) growth is depicted. This network has 64 inputs and 64 outputs, which better describes living systems, it is not autonomous network as in ch.3. As mentioned above, chaos is a  high probability of large damage avalanche after small disturbance. Damage $d$ (horizontal axis) is measured as a fraction of all $N$ nodes which have different state than in the not disturbed network. Small network where $N=50$ behave like at the edge of chaos and order. Its distribution is near power law, as for the avalanches in the pile of sand which Per Bak's self-organized criticality \citep{Bak96} controls. But living objects have a method against such control - it is self-multiplying. In the effect typically a part of objects avoid large avalanche and can grow. This way they enter chaotic regime and can stay there. Now distribution consists of two peaks: The right one stay in the equilibrium point after large avalanche. Such large avalanches occur with high probability which define system as chaotic. Here `matured chaos' starts about $N=600$, better $N=1000$ where very low probability occurs between peaks, which I discuss in Ref. \citep{bgec}. Large avalanches change most of node states which means that after such avalanche system works in  totally different way. For living objects it means death, i.e. elimination. It cannot be a new evolutionary change (compare to \citep{Farmer86}) - it is much more than even Lysenko proposed.  Can you imagine that more than half of all mechanisms changed and living object survived? 

But there remains left peak which is similar to that for $N=50$. Now, however, with a real level of probability it covers only a small fraction of available nodes (each time a different set) but it is enough for adaptive evolution and life, which use only small changes. Without right peak there is no qualitative difference between new state of still living object and dead object. It is possible to investigate such simple model without such difference and with power law distribution, but it is huge simplification. This qualitative difference is important to differentiate between elimination and surviving, which is a base of Darwinian mechanism. Therefore edge of chaos with one left peak of power law distribution cannot be adequate area for living objects on which Darwinian mechanism works.
When network grows, or its other parameters change causing growth of chaos level and decreasing order level, then probability of large avalanche also grows and probability of surviving decreases. Therefore life cannot quickly run deep into chaos, however, Darwinian mechanism collects certain mechanisms which can neutralize certain types of initialising changes. Now these known type (ch.1.3.1.) initialisations does not lead to large avalanche, i.e. elimination, and chance of survival goes up.  
Kauffman (and I too) does not model these `known type initialisations' nor connected to them homeostatic answers of system, based typically on negative feedbacks, but this large simplification must be taken under consideration when stability of living objects is discussed. Authors of Ref. \citep{Wagner01,Serra04,Ramo06,Serrajtb07} compare real biological stability measured by \citet{Hughes00}  to structural stability near the edge of chaos and they obtain similarity. They treat this similarity as the main evidence that living objects stay on the edge of chaos but they neglect taking this simplification into account.

\begin{figure}
\begin{center}
\includegraphics*[width=8.5cm]{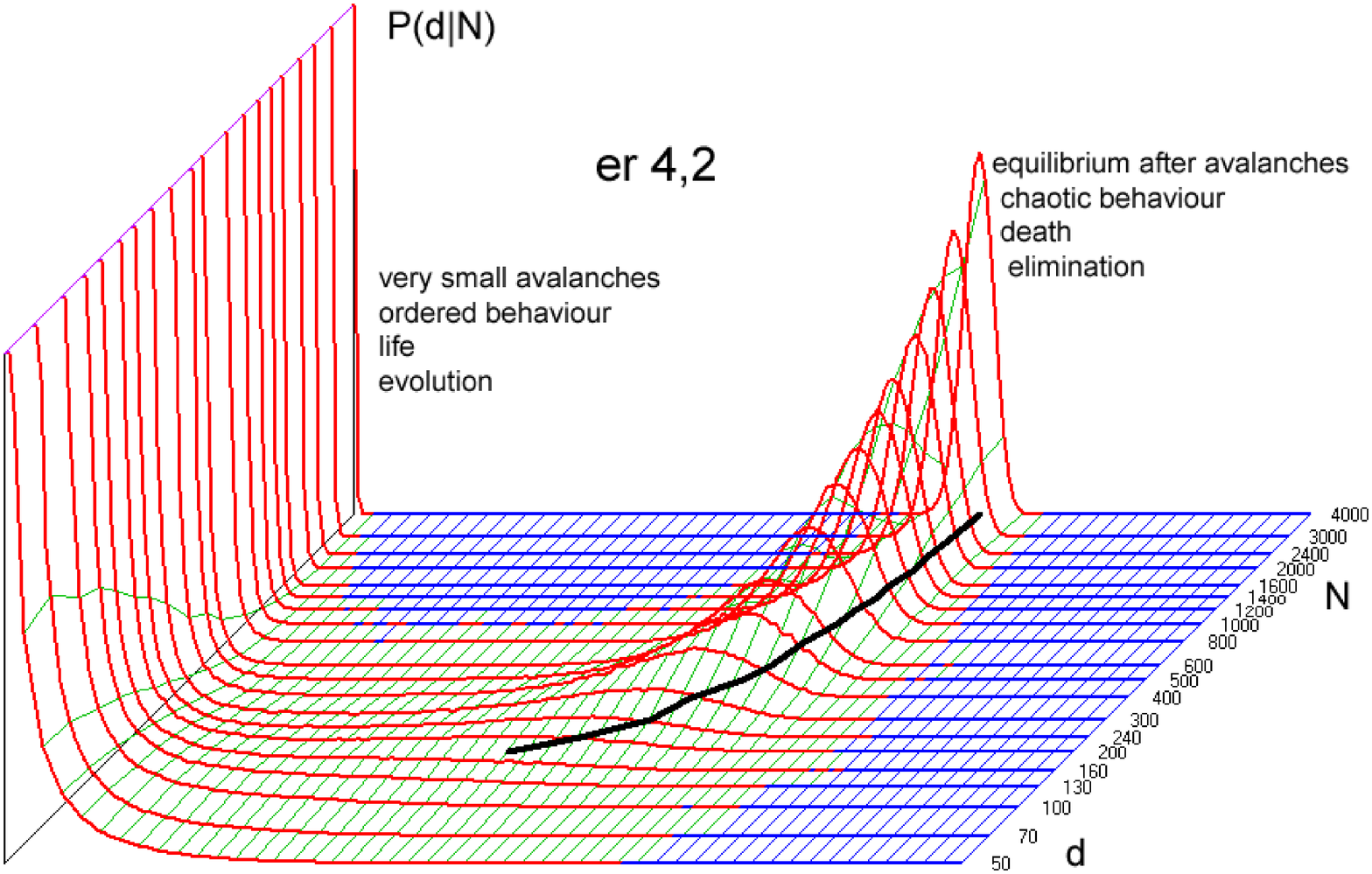}
\end{center}

\caption{Damage size $d$ distribution during network $er$ 4,2 (64 inputs and outputs) growth. Emergence of chaos and death (right peak with large avalanche). Damage $d$ is measured as a fraction of all $N$ nodes which have different state than in the no disturbed network. Small network ($N=50$) behave like on the edge of chaos and order (near power law). This peak approximately remains as left one in chaotic area and there is a place for life evolution.}

\label{fig:8}     
\end{figure}



\end{document}